\documentclass[twocolumn]{aa} 
\usepackage{graphicx}
\usepackage{subfigure}
\usepackage{txfonts}
\begin{document}
   \title{The Ara OB1a association}
   \titlerunning{Ara OB1a region}

   \subtitle{Stellar population and star formation history}

   \author{G. Baume\inst{1}\fnmsep\thanks{Member of Carrera del Investigador CONICET, Argentina},
           G. Carraro\inst{2}\fnmsep\thanks{On leave from Dipartimento di Astronomia,
           Universit\`a di Padova, Vicolo Osservatorio 2, I-35122, Padova, Italy},
           F. Comeron\inst{3},
          \and
           G. C. de El\'ia\inst{1}
          }
   \authorrunning{Baume et al.}

   \institute{Facultad de Ciencias Astron\'omicas y Geof\'{\i}sicas (UNLP),
              Instituto de Astrof\'{\i}sica de La Plata (CONICET, UNLP),
              Paseo del Bosque s/n, La Plata, Argentina \\
              \email{gbaume@fcaglp.fcaglp.unlp.edu.ar} \\
              \email{gdeelia@fcaglp.fcaglp.unlp.edu.ar}
   \and
              ESO, Alonso de Cordova 3107, Vitacura, Santiago de Chile, Chile \\
              \email{gcarraro@eso.org}
   \and
              ESO, Karl-Schwarzschild-Strasse 2 85748 Garching bei Munchen Germany  \\
              \email{fcomeron@eso.org}
             }

   \date{Received: June 10, 2010; Accepted: April 26, 2011}


  \abstract
   {The Ara OB1a association is a nearby complex in the fourth Galactic quadrant where a 
    number of young/embedded star clusters are projected close to more evolved, intermediate 
    age clusters. It is also rich in interstellar matter, and contains evidence of the 
    interplay between massive stars and their surrounding medium, such as the rim HII region 
    NGC 6188.}
   {We provide robust estimates of the fundamental parameters (age and distance) of the two 
    most prominent stellar clusters, NGC~6167 and  NGC~6193, that may be used as a basis for 
    studing the star formation history of the region.}
   {The study is  based on a photometric optical survey ($UBVIH_{\alpha}$) of NGC~6167 and
    NGC~6193 and their nearby field, complemented with public data from 2MASS-VVV, UCAC3, 
    and IRAC-Spitzer in this region.}
   {We produce a uniform photometric catalogue and estimate more robustly the fundamental  
    parameters of NGC~6167 and NGC~6193, in addition to the IRAS~16375-4854 source. As a 
    consequence, all of them are located at approximately the same distance from the Sun 
    in the Sagittarius-Carina Galactic arm. However, the ages we estimate differ widely: 
    NGC~6167 is found to be an intermediate-age cluster (20-30 Myr), NGC~6193 a very young 
    one (1-5 Myr) with PMS, $H_{\alpha}$ emitters and class II objects, and the 
    IRAS~16375-4854 source is the youngest of the three containing several YSOs.}
   {These results support a picture in which Ara OB1a is a region where star formation has 
    proceeded for several tens of Myr until the present. The difference in the ages of 
    the different stellar groups can be interpreted as a consequence of a triggered star 
    formation process. In the specific case of NGC~6193, we find evidence of possible 
    non-coeval star formation.}


   \maketitle

\section{Introduction}

\begin{figure*}
\begin{center}
\includegraphics[width=15cm]{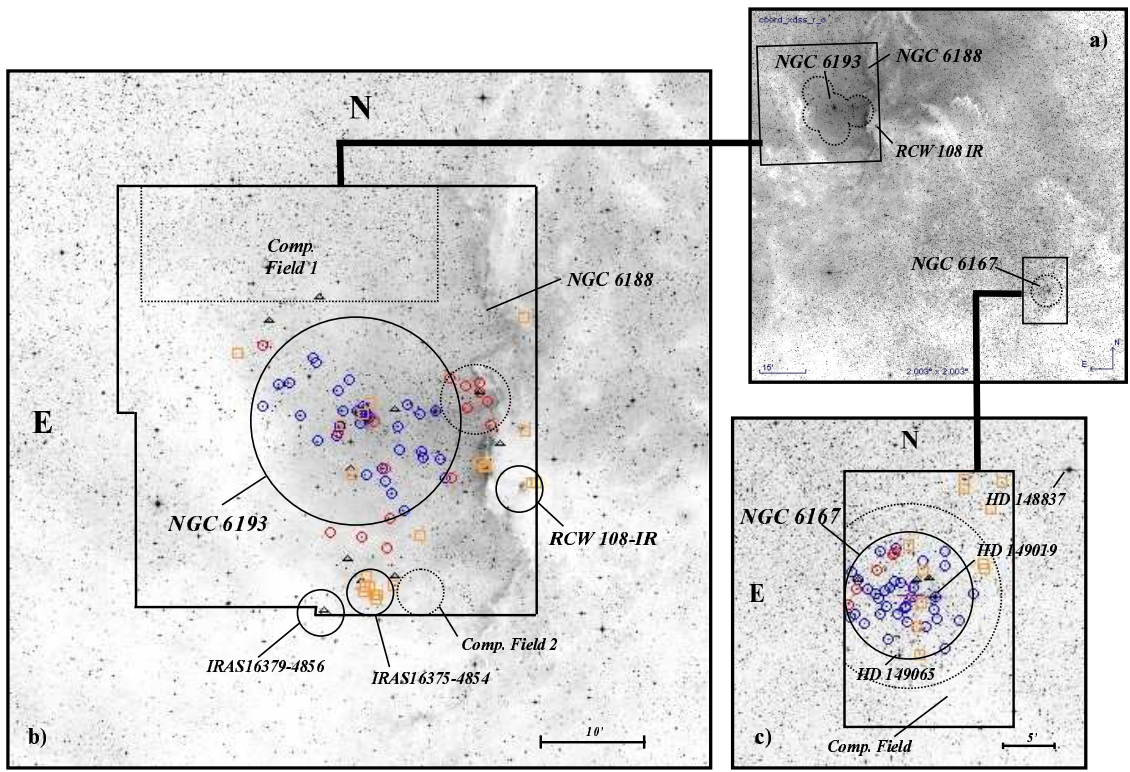}
\caption{Second generation Digitized Sky Survey (DSS-2; red  filter) images. {\bf a)} Image 
centered at $\alpha_{J2000} = 16:38:00$; $\delta_{J2000} \sim -49:15:20$ and covering a 
$2^\circ \times 2^\circ$ field. Solid rectangles centered in NGC~6167 and NGC~6193 indicates 
areas surveyed with $UBVI$ observations. Dotted circle around NGC~6167 and several dotted 
circles around NGC~6193 indicates the areas with $H_{\alpha(on/off)}$ data. {\bf b)} and 
{\bf c)} Detailed images centered respectively at NGC~6193 and NGC~6167 showing main objects 
in each region and the adopted comparison fields. For the meaning of the symbols see in 
advance Fig~\ref{fig:tcds1} caption.}
\label{fig:dss}
\end{center}
\end{figure*}

Ara OB1a has been suggested as an example of triggered star formation in the local Galaxy.
A summary of the history of investigations of this region, our present understanding, and
possible future studies is presented in Wolk et al. (2008a,b and references therein). This 
nearby association contains three open clusters (NGC~6193 in the center of the association, 
together with NGC~6167 and NGC~6204). In addition, several embedded clusters are present 
together with both star-forming and quiescent clouds (see Fig.~1 in Wolk et al 2008a and 
Fig.~\ref{fig:dss} in this paper). The idea that triggered star fromation has taken place in 
Ara OB1a is suggested by a variety of observations. The prominent cluster NGC~6193 appears 
to be connected to RCW 108 (=NGC~6188), a rim HII region located westward that marks the 
edge of a molecular cloud being eroded by the hottest stars in the cluster, where several 
IRAS sources have been found (Comer\'on et al. 2005). Star formation in the molecular cloud 
may have been triggered by NGC 6193. A connection between the energetic activity of the 
stars in NGC 6167 and a giant HI bubble surrounding was proposed by Arnal et al. (1987) and 
Waldhausen et al. (1999) (see Fig.~6 at Wolk et al. 2008b).
 
The ages and distances of the two major clusters NGC~6167 and NGC~6193 derived thus far have 
been based on the photoelectric and photographic photometry of a few stars, rather than on 
modern CCD data. Carraro \& Munari (2004) presented CCD data for NGC~6204, but this cluster 
is located in the outskirts of the association and may not be related to the main 
star/cluster formation process. Distance estimates to NGC~6193 range from $\sim$ 1100 pc to 
$\sim$ 1400 pc, and its age is $\sim$ 3 Myr (Moffat \& Vogt 1973; Fitzgerald 1987; Kaltcheva 
\& Georgiev 1992; V\'azquez \& Feinstein 1992). As for NGC~6167, the estimated distance 
ranges from 600 to 1200 pc and the age from 10 Myr to 40 Myr (Moffat \& Vogt 1975; Bruck 
\& Smyth 1967; Rizzo \& Bajaja 1994).

In the present paper, we present a new CCD-based optical ($UBVI_C$ and $H_\alpha$) survey in 
an attempt to provide more robust determinations of the age and distance of NGC~6193 and 
NGC~6167. The photometry we present is sufficiently deep to allow us to estimate the nuclear 
and contraction age, by detecting pre-main sequence (PMS) stars with $H_\alpha$ emission. 
Therefore we expect to provide firmer estimates of the age and at the same time highlight 
possible differences between nuclear and contraction ages.

The layout of the paper is as follows. In Sect.~\ref{sec:data}, we describe the data sources,
and their reduction and calibration. In Sect.~\ref{sec:analysis}, we present the analysis of
the data together with the derivation of the main clusters parameters. In 
Sect.~\ref{sec:objects}, we derive the main properties of NGC~6167, NGC~6193, and the 
IRAS 16375-4854 source. Finally, in Sect.~\ref{sec:discussion} we discuss and summarize our 
results.

\section{Data} \label{sec:data}

\subsection{Optical data}

Our main data set is based on CCD photometric observations of stars in the region of 
NGC~6167 and NGC~6193 (see Fig.~\ref{fig:dss}) carried out during several observational runs, 
and is complemented with available information from the literature. These data include:

\begin{itemize}
\item Las Campanas Observatory (LCO) $UBVI_C$ images of the region of NGC~6167 acquired
   during two runs (25 June 2006 and 24 June 2009) using the 1.0 m Swope telescope equipped 
   with the Site 3 $2048 \times 3150$ CCD camera. The field of view (FOV) of these images is 
   about $14\farcm8 \times 22\farcm8$ with a scale $0\farcs435$/pix. The typical FWHM of the 
   data was about $0\farcs9$ and airmas values during the observation of the scientific 
   frames were in the range 1.07-1.12.
\item Cerro Tololo Inter-American Observatory (CTIO) $UBVI_C$ images of the region covering 
   about $40\farcm0 \times 40\farcm0$ around NGC~6193 acquired during two runs (two frames 
   on 25 and 26 March 2006 and two frames on 21 May 2006). We used the Y4KCAM camera 
   attached to the 1.0m telescope, which is operated by the SMARTS 
   consortium\footnote{http://www.astro.yale.edu/smarts}. 
   This camera is equipped with an STA $4064 \times 4064$ CCD with $15 \mu$ pixels. This  
   set-up provides direct imaging over a FOV $20\farcm0 \times 20\farcm0$ with a scale of 
   $0\farcs289$/pix\footnote{http://www.astronomy.ohio-state.edu/Y4KCam/detector.html}.
   Typical FWHM of the data was about $0\farcs9$, and airmasses during the observation of 
   the scientific frames were in the range 1.02-1.28 .
\item Complejo Astron\'omico El Leoncito (CASLEO) $H_{\alpha-on}$ (656.6 nm) and 
   $H_{\alpha-off}$ (666.6 nm) observations.  They were performed on 14, 15, 17 and 18 May 
   2002 in six circular frames ($4\farcm5$ radius) covering the area of NGC~6167 (1 frame) 
   and NGC~6193 (5 frames). We used a nitrogen-cooled detector Tek-1024 CCD and focal 
   reducer attached to the 215-cm telescope (scale $0\farcs813$/pix). Exposure times in 
   $H_{\alpha-on}$ and $H_{\alpha-off}$ ranged from $2~s$ to $150~s$.

\item Other data sources: photometric data for a few bright stars that were found to be 
   saturated in our observations, were taken from Webda 
   database\footnote{http://www.univie.ac.at/webda}.
\end{itemize}

Details of all these observations are given in Table~\ref{tab:frames}.

\begin{table}
\begin{center}
\fontsize{8} {10pt}\selectfont
\caption{Journal of observations of the scientific frames together with
the derived photometric calibration coefficients.}
\begin{tabular}{clccccc}
\hline
\multicolumn{7}{l}{Exposure times [sec]} \\
\hline
Observ. & \multicolumn{1}{l}{Frames} & $U$ & $B$ & $V$ & $I_C$ & $N$ \\
\hline
LCO   & long   & ~~~-~~~     & ~~~900      & ~600 & ~600 & 1 \\
      & short  & 4$\times30$ & 2$\times$20 & ~~10 & ~~10 & 1 \\
\hline
CTIO  & long   & 1500 & 1200 & ~900 & ~700 & 4 \\
      & medium & ~200 & ~100 & ~100 & ~100 & 4 \\
      & short  & ~~30 & ~~30 & ~~30 & ~~30 & 4 \\
      & vshort & ~~10 & ~~~7 & ~~~5 & ~~~5 & 4 \\
\hline
\hline
\multicolumn{7}{l}{Calibration and extinction coefficients} \\
\hline
Coefficient & \multicolumn {3}{c}{LCO}          & \multicolumn {3}{c}{CTIO}               \\
\hline
$u_1$ & \multicolumn {3}{c}{$+4.528 \pm 0.006$} & \multicolumn {3}{c}{$+3.254 \pm 0.007$} \\
$u_2$ & \multicolumn {3}{c}{$+0.125 \pm 0.011$} & \multicolumn {3}{c}{$-0.034 \pm 0.013$} \\
$u_3$ & \multicolumn {3}{c}{$+0.490 \pm 0.010$} & \multicolumn {3}{c}{$+0.450$}           \\
\hline
$b_1$ & \multicolumn {3}{c}{$+2.938 \pm 0.003$} & \multicolumn {3}{c}{$+2.015 \pm 0.021$} \\
$b_2$ & \multicolumn {3}{c}{$+0.050 \pm 0.004$} & \multicolumn {3}{c}{$+0.120 \pm 0.025$} \\
$b_3$ & \multicolumn {3}{c}{$+0.250 \pm 0.010$} & \multicolumn {3}{c}{$+0.250$}           \\
\hline
$v_1$ & \multicolumn {3}{c}{$+2.864 \pm 0.003$} & \multicolumn {3}{c}{$+1.802 \pm 0.013$} \\
$v_2$ & \multicolumn {3}{c}{$-0.068 \pm 0.004$} & \multicolumn {3}{c}{$-0.048 \pm 0.016$} \\
$v_3$ & \multicolumn {3}{c}{$+0.160 \pm 0.010$} & \multicolumn {3}{c}{$+0.160$}           \\
\hline
$i_1$ & \multicolumn {3}{c}{$+3.757 \pm 0.007$} & \multicolumn {3}{c}{$+2.676 \pm 0.020$} \\
$i_2$ & \multicolumn {3}{c}{$+0.039 \pm 0.009$} & \multicolumn {3}{c}{$-0.019 \pm 0.021$} \\
$i_3$ & \multicolumn {3}{c}{$+0.080 \pm 0.010$} & \multicolumn {3}{c}{$+0.080$}           \\
\hline
\label{tab:frames}
\end{tabular}
\end{center}
\tablefoot{$N$ indicates the number of fields observed for  each
cluster region.}
\end{table}

\begin{table}
\begin{center}
\fontsize{8} {10pt}\selectfont
\caption{Approximate $V$ magnitud limit values for 80\% and 50\% completeness factors for each studied band.}
\begin{tabular}{ccccc}
\hline
Filter       & \multicolumn{2}{c}{NGC~6167} & \multicolumn{2}{c}{NGC~6193}  \\
             &  80\%    &   50\%            &  80\%        &  50\%          \\
\hline
$U$          &  13      &   14              &  16      &   17               \\
$B$          &  20      &   21              &  18      &   19               \\
$V$          &  21      &   23              &  21      &   23               \\
$I$          &  21      &   23              &  21      &   23               \\
$H_{\alpha}$ &  13      &   15              &  13      &   15               \\
$J$          &  16      &   18              &  17      &   19               \\
$H$          &  16      &   18              &  17      &   19               \\
$K$          &  16      &   18              &  17      &   19               \\
$[3.6]$      &  12      &   15              &  13      &   16               \\
$[4.5]$      &  12      &   15              &  13      &   16               \\
$[5.8]$      &  12      &   15              &  13      &   16               \\
$[8.0]$      &  12      &   15              &  13      &   16               \\
\hline
\label{tab:comp}
\end{tabular}
\end{center}
\end{table}

\subsubsection{Reduction}

All frames were pre-processed in a standard way using the IRAF\footnote{IRAF is distributed 
by NOAO, which is operated by AURA under cooperative agreement with the NSF.} package CCDRED. 
For this purpose, zero exposures, and sky flats were taken every night. Photometry was 
performed using IRAF DAOPHOT and PHOTCAL packages. Instrumental magnitudes were obtained 
using the point spread function (PSF) method (Stetson 1987). Since the FOV is large, a 
quadratic spatially variable PSF was adopted and its calibration for each image performed 
using several isolated, spatially well distributed, bright stars (about 25) across each 
field. The PSF photometry was aperture-corrected for each filter and exposure time. 
Aperture corrections were computed by performing aperture photometry of a suitable number 
(about 20) of bright stars in the field. Finally, all data from different filters and  
exposures were combined and calibrated using DAOMASTER (Stetson 1992).

\subsubsection{Photometric calibration} \label{sec:calib}

Standard stars were selected from particular areas of the catalog of Landolt (1992) (Mark A, 
PG 1323, and SA 110 for LCO data and SA 101, SA 104, and SA 107 for CTIO data) were used to 
determine the transformation equations relating our instrumental magnitudes to the standard 
$UBVI_C$ system. The selection of the fields was made to ensure that the stars had a wide 
range in colours. Aperture photometry was then carried out for all the standard stars 
($\sim 70$ per night) using the IRAF PHOTCAL package. To tie our observations to the 
standard system, we use transformation equations of the form\\

\begin{center}
\begin{tabular}{llc}
$u = U + u_1 + u_2 (U-B) + u_3 X$           & ($rms = 0.04$) & (1), \\
$b = B + b_1 + b_2 (B-V) + b_3 X$           & ($rms = 0.03$) & (2), \\
$v = V + v_1 + v_2 (B-V) + v_3 X$           & ($rms = 0.02$) & (3), \\
$i = I_C + i_1 + i_2 (V-I_C) + i_3 X$       & ($rms = 0.02$) & (4), \\
\end{tabular}
\end{center}

\noindent where $UBVI_C$ and $ubvi$ are standard and instrumental magnitudes respectively 
and $X$ is the airmass of the observation. The used transformation coefficients and 
extinction coefficients are shown at the bottom of Table~\ref{tab:frames}.

The $H_{\alpha-on/off}$ observations were calibrated by fitting the empirical MS (Didelon 
1982 and Feigelson 1983) to the adopted MS stars in each cluster field (see 
Sect.~\ref{sec:halpha} and Fig.~\ref{fig:cmdha}). In this way, the $H_{\alpha}(on) - 
H_{\alpha}(off)$ index (hereafter $\Delta H_{\alpha}$) has a negative value for objects with 
a $H_{\alpha}$ line above the continuum. The relation between the $\Delta H_{\alpha}$ index 
and the $H_{\alpha}$ line width ($W_{H\alpha}$) is given by 
$\Delta H_{\alpha} = -2.5~log (1 - W_{H\alpha} [\AA] / 60)$.

\begin{figure}
\begin{center}
\includegraphics[width=6cm]{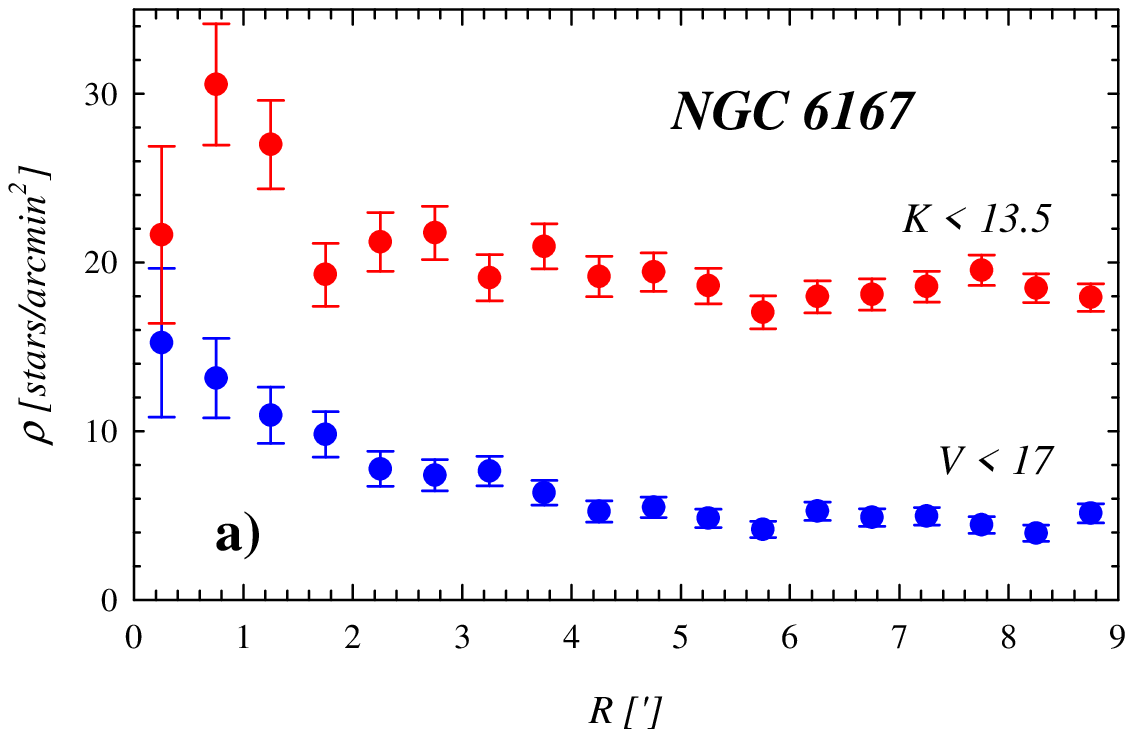}
\includegraphics[width=6cm]{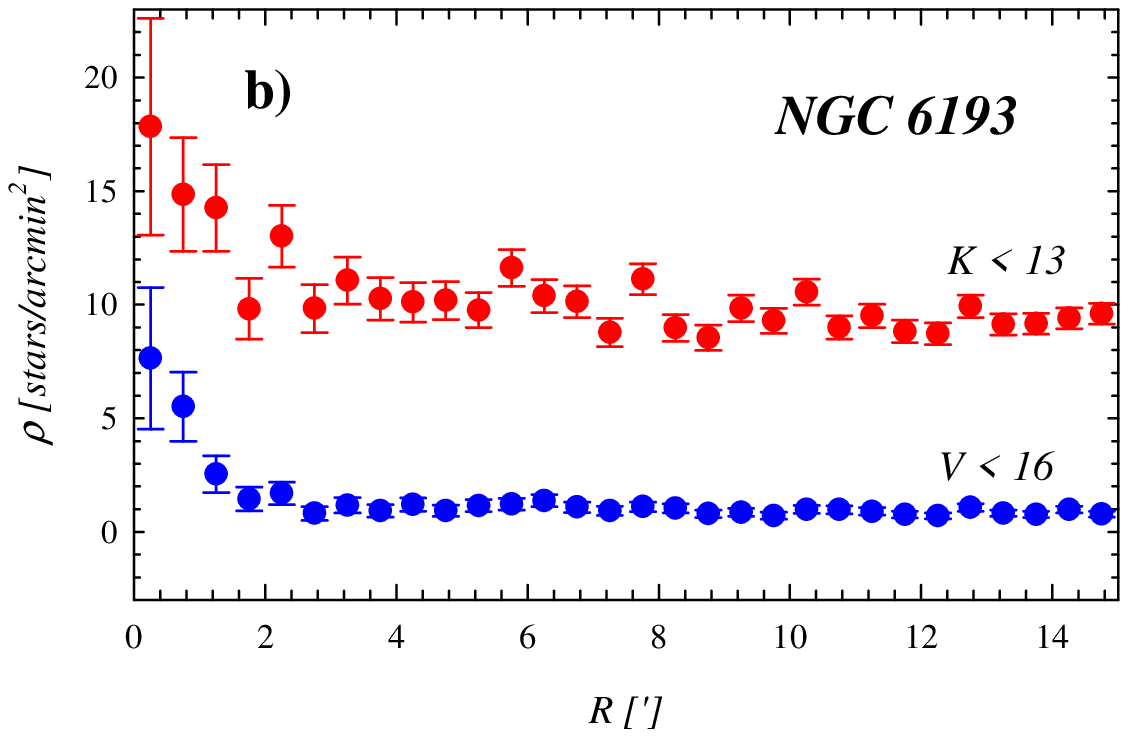}
\caption{Radial density profiles for NGC~6167 (upper panel) and NGC~6193 (lower panel).
Lower plots (blue) correspond to CCD data ($V < 16$), whereas upper plots (red) correspond
to 2MASS data ($K < 13$). Poisson error bars ($1\sigma$) are also shown.}
\label{fig:radial}
\end{center}
\end{figure}

\begin{figure}
\centering
\subfigure{\includegraphics[height=4.5cm]{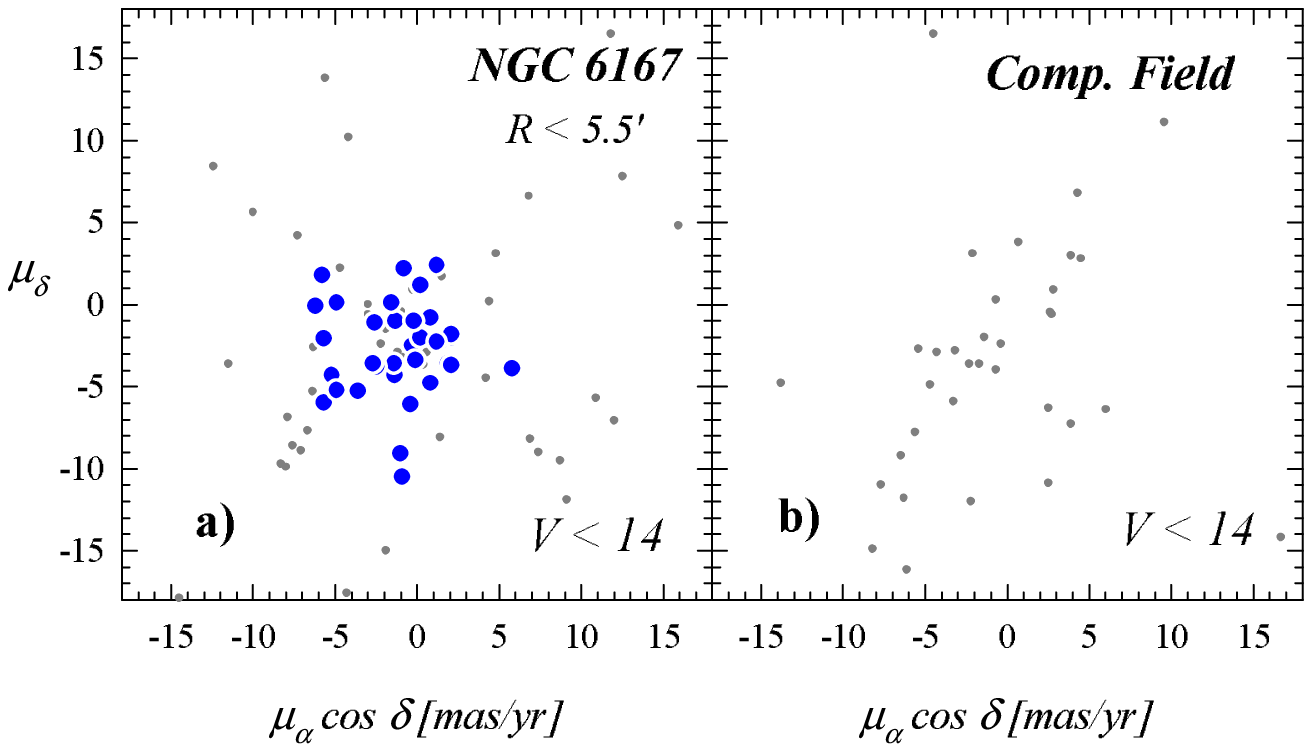}}
\subfigure{\includegraphics[height=4.5cm]{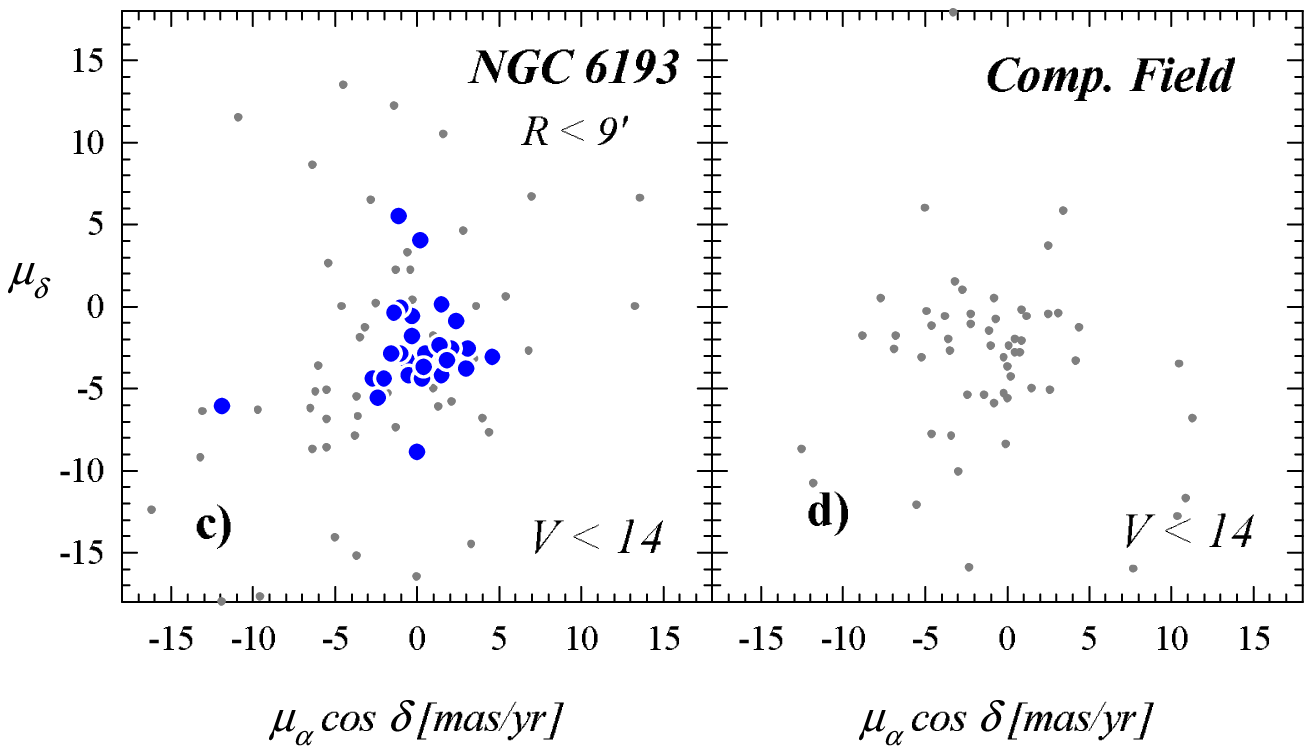}}
\caption{Proper motion VPDs for adopted cluster areas of NGC~6167 and NGC~6193 and their corresponding
  comparison fields. Filled (blue) circles are adopted as likely members for each cluster.}
\label{fig:vpds}
\end{figure}

\subsection{Mid-infrared data}

The NGC~6193 and NGC~6167 regions were observed in the mid-IR by the Infrared Array Camera 
(IRAC) on-board the Spitzer Space Telescope (SST). IRAC was used to obtain images in four 
channels (3.6, 4.5, 5,8 and 8.0 $\mu m$) and a catalogue was produced using several of these 
photometric data points (GLIMPSE\footnote{http://www.astro.wisc.edu/glimpse/docs.html}). 
Observations covering NGC~6167 region were included in this catalogue. However, in the case 
of NGC~6193 and its surrounding area, only IRAC images were available in the SST database
belonging to programs P00112, P00191, P03536, P20597, P30570, and P40321.

We performed aperture photometry in all these images by using the IRAF PHOTCAL package 
following the procedure described by the Spitzer Science 
Center\footnote{http://ssc.spitzer.caltech.edu/irac/}. 
Briefly, we run DAOFIND task to look for point sources. We then used the PHOT task using an 
aperture of 5 pixels and a background annulus from 12 to 20 pixels. Finally, to calibrate 
our results, we adopted the aperture correction values given in the on-line calibration 
tables. To avoid false detections, we selected only sources detected in both the four IRAC 
channels and the $V$ band images.

\subsection{Complementary data and astrometry}

Using the X--Y stellar positions obtained from our data, we correlated their positions with 
data from the following catalogues: a) ``Two-Micron All Sky Survey`` (2MASS; Cutri et al. 
2003, Skrutskie et al. 2006); b) ``Vista Variables in the V\'ia L\'actea'' catalogue 
performed at ``Cambridge Astronomy Survey 
Unit''\footnote{http://casu.ast.cam.ac.uk/vistasp/vvv} 
(VVV+CASU; Minniti et al. 2010), and c) ``The Third U.S. Naval Observatory CCD Astrograph 
Catalog'' (UCAC3; Zacharias et al. 2010).

Our adopted procedure to perform the astrometric calibration of our data was explained in 
Baume et al. (2009). The rms of the residuals in the positions were $\sim 0\farcs17$, which 
is about the astrometric precision of the 2MASS catalogue ($\sim0\farcs12$). As for NIR 
photometry, only stars detected in $V$ filter were selected from NIR catalogues. We adopted 
2MASS values only for the brightest stars ($K < 13$) and VVV+CASU ones for the remainder.

\subsection{Data completeness}

The completeness of the observed star counts is a relevant issue, thus it was computed by 
means of artificial-star experiments on our data (see Carraro et al. 2005). We created 
several artificial images by adding at random positions a total of 20 000 artificial stars 
to our true images. These were distributed with a uniform probability distribution of the 
same colour and luminosity as the real sample. To avoid overcrowding, in each experiment we 
added the equivalent to only 15\% of the original number of stars. Since in all bands were 
considered only detections with a $V$ filter counterpart, we performed those experiments on 
the long exposure image of $V$ filter computing the completeness factor as the ratio of the 
number of artificial stars recovered to the number of artificial stars added. The 
corresponding factors for the other bands were computed in a relative way as the ratio of
the amount of detected stars in each band to the amount of detected stars in $V$ filter 
corrected by completeness (weighted by diferent area coverage if necessary).

The computed values of the completeness factor for different $V$ magnitude bins are listed 
in Table~\ref{tab:comp}. We note that for NGC~6167, only short exposures were obtained in 
the $U$ filter, which explains the poor completeness of its catalogue in this band. 
Notwithstanding the completeness is sufficient to ensure a reliable estimate of the cluster 
parameters.

\subsection{Final catalogues}

The above procedure allowed us to build two astrometric, photometric (13 bands) catalogues 
covering the regions of both NGC~6167 (33193 objects) and NGC~6193 (39338 objects). These 
two catalogues constitute the main observational database used in this study. A solid 
analysis of the behavior of the stellar energy distributions (SEDs) can be carried out with 
this tool, thus preventing possible degeneracies in the photometric diagrams and allowing us 
to obtain more reliable results. Both catalogues are available only in electronic form at 
the Centre de Donnes astronomiques de Strasbourg (CDS). They includes X--Y positions, 2MASS 
identification (when available), equatorial coordinates (epoch 2000.0), and optical ($UBVI$ 
and $H_{\alpha-on/off}$), NIR 2MASS/VVV ($JHK$), and MIR IRAC-SST (3.6, 4.5, 5,8 and 8.0 
$\mu m$) photometry.

\section{Data analysis} \label{sec:analysis}

\begin{figure*}
\begin{center}
\includegraphics[width=13cm]{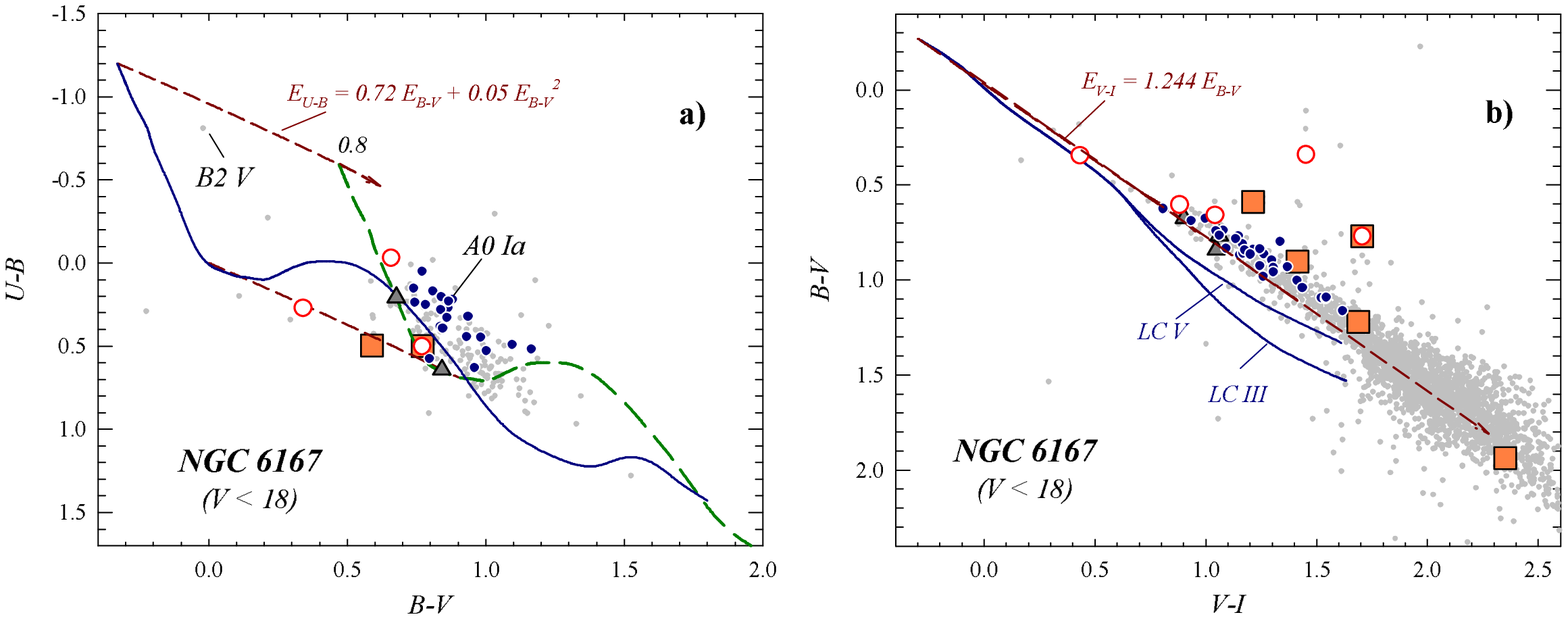}
\includegraphics[width=13cm]{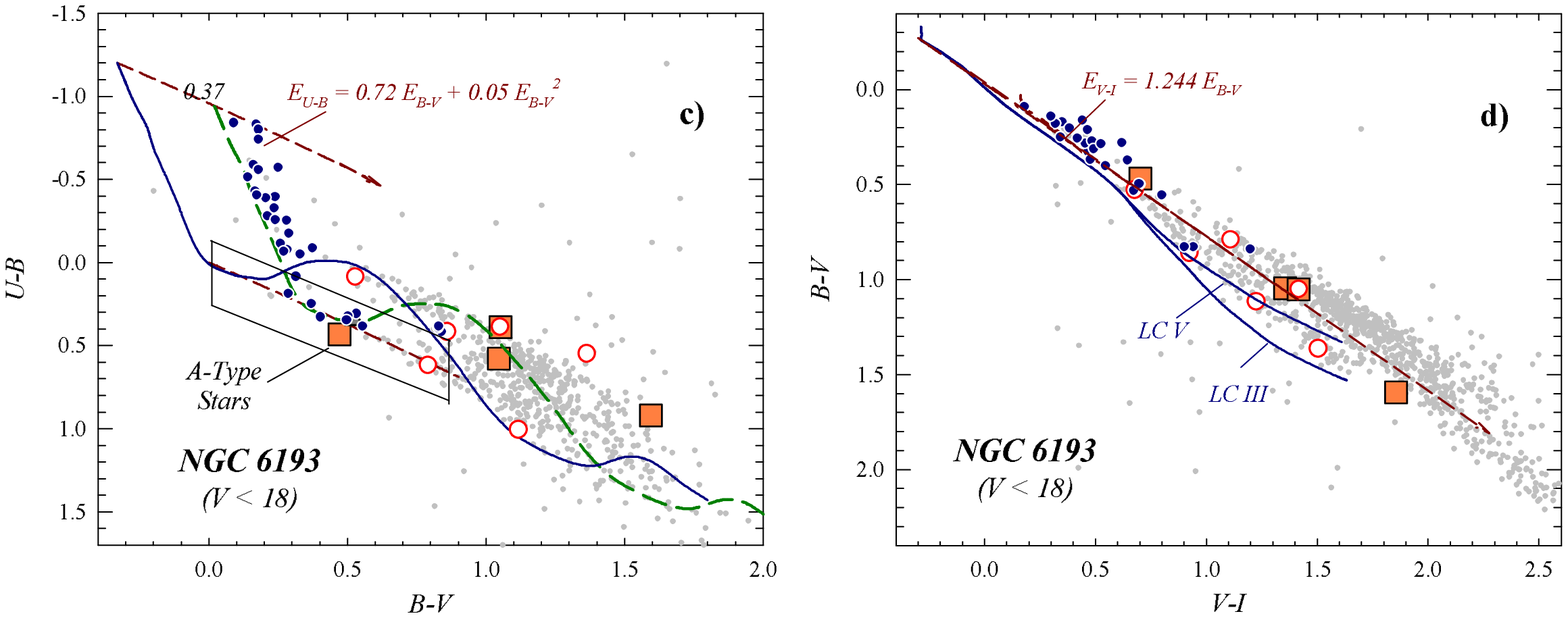}
\caption{Optical TCDs of stars located in NGC~6167 (upper panel) and NGC~6193 (lower panel) 
  regions. $U-B$ vs. $B-V$ diagrams: filled (blue) circles are adopted likely members for 
  each cluster, open (red) circles are likely $H_{\alpha}$ emitter stars; big (orange) 
  squares and filled (grey) triangles are respectively considered probable class I and class 
  II objects (see Sect~\ref{sec:ir} and also the comments for each cluster); light dots are 
  considered as field stars. The solid (blue) curve is the Schmidt-Kaler (1982) ZAMS, while 
  dashed (green) lines are the same ZAMS, but shifted along the reddening line (red) by the 
  adopted colour excesses indicated above them. They correspond to the adopted values for 
  the cluster stars (see Table~\ref{tab:param}). $B-V$  vs. $V-I$ diagrams: Symbols have the 
  same meaning as in plots a) and c), whereas solid curves (blue) are intrinsic colours for
  luminosity class V and III from  Cousins (1978ab). Dashed (red) arrows indicate the normal 
  reddening path ($R_V = 3.1$).}
\label{fig:tcds1}
\end{center}
\end{figure*}

\subsection{Clusters centers and sizes}

Radial stellar density profiles provide an objective tool to determine the size of a star 
cluster. They were computed, in a first attempt, starting from  the estimated position of 
the cluster centers obtained by visual inspection of the second generation Digitized Sky 
Survey (DSS-2, red) plates, using our brightest catalogued stars ($V < 16-17$ depending on 
the cluster) and all the brightest 2MASS catalogued objects ($K < 13-13.5$ depending on the 
cluster). We then repeated the procedure several times, slightly changing in each case the 
position of the centers and using different magnitude bins. We finally adopted these cases 
that produce the most uniform fall off in both bands. According to the variation observed in 
different experiments, we estimated that the clusters centers have $\sim 0\farcm5$ 
uncertainty in each coordinate.

The cluster radial density profiles were computed in the usual way (see Baume et al. 2004), 
namely calculating the surface density in concentric, equal-area rings. Since we applied 
this method to both the optical ($V$ filter) and the 2MASS infrared data ($K$ filter), this 
allowed us to detect the possible influence of interstellar material. In both clusters, 
noticeable differences between the profiles for each band are detected. As a result, the 
sizes were estimated taking into account mainly the $K$ filter, where absorption effects 
are minimized.

The adopted centers and radii are indicated in Table~\ref{tab:param} and Fig.~\ref{fig:dss}. 
Centers are close to the ones given by Dias et al. (2002) or by the $SIMBAD$ database, 
although the radius values are larger than those compiled by Dias et al. (2002), which are 
based only on visual inspection. The radial density profiles are shown in 
Fig.~\ref{fig:radial}. Our data completely cover the clusters and sample also a significant 
portion of the surrounding field. This allows us to select different zones in each region 
and study them separately to understand the behavior of the stellar population in the 
directions of the clusters and their neighborhood (see Fig.~\ref{fig:dss}).

\begin{figure*}
\begin{center}
\includegraphics[width=12cm]{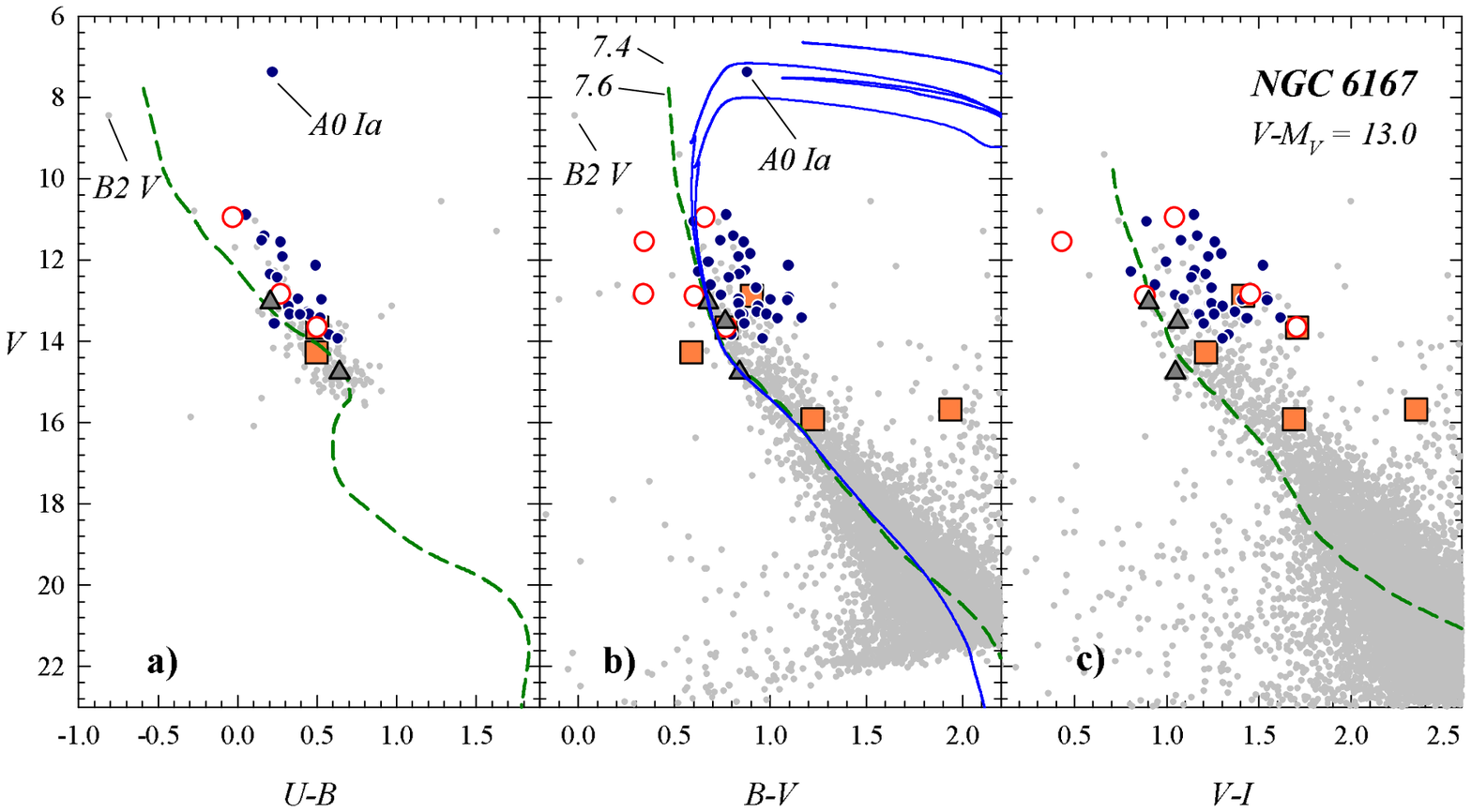}
\includegraphics[width=12cm]{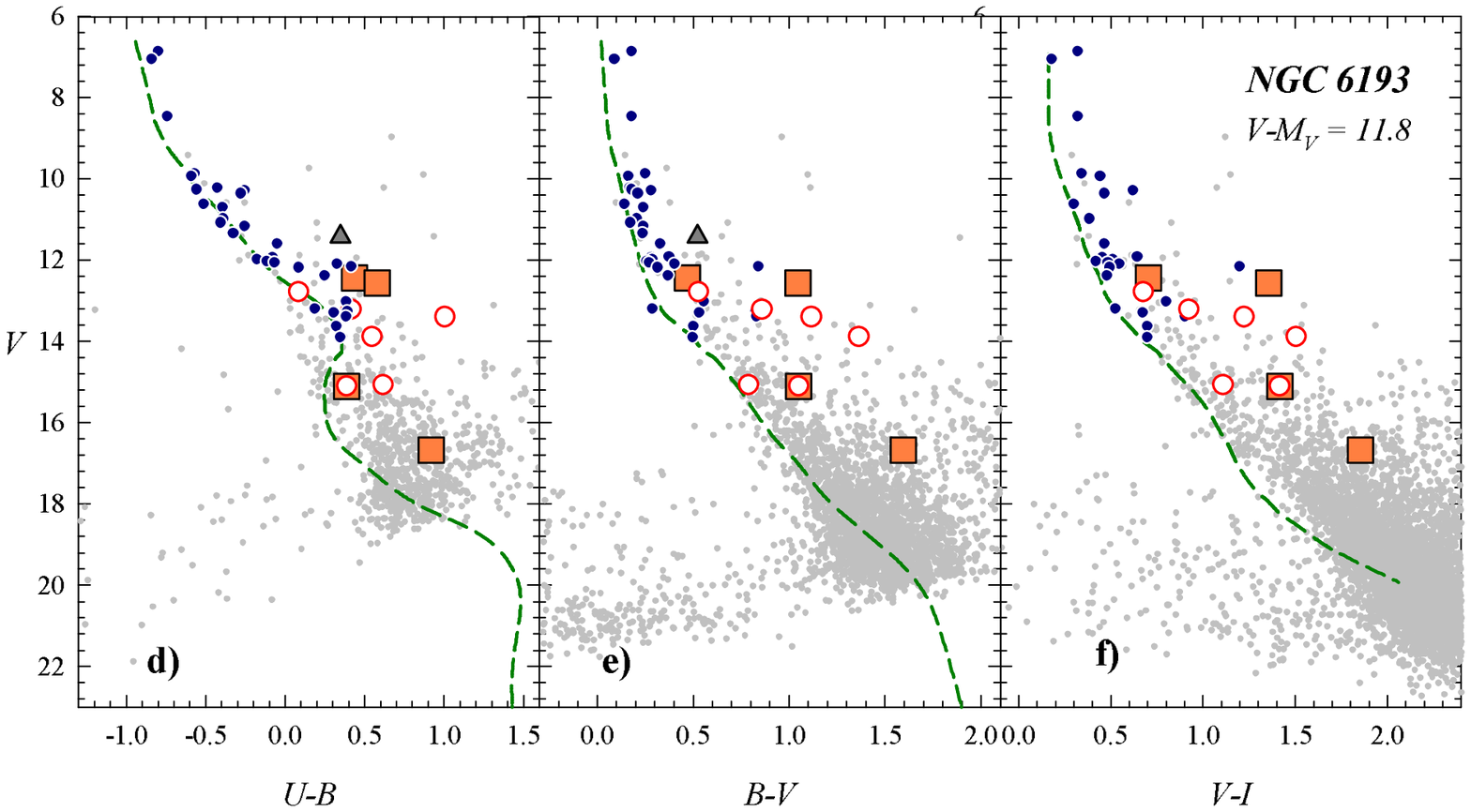}
\caption{Optical CMDs of stars located in NGC~6167 (upper panels) and NGC~6193 (lower panels) regions.
  Symbols are the same as in Fig.~\ref{fig:tcds1}. Dashed (green) curves are the Schmidt-Kaler (1982)
  empirical ZAMS and the MS path from Cousins (1978ab). Solid curves are Marigo et al. (2008) isochones.
  All the reference curves are corrected by the adopted colour excess and apparent distance modulus for
  each cluster (see Table~\ref{tab:param}).}
\label{fig:cmds1a}
\end{center}
\end{figure*}

\subsection{Proper motion analysis} \label{sec:pmotion}

To obtain additional information to confirm and support the membership assignment, we used
kinematic data provided by UCAC3 and Hipparcos catalogues (the later one was necessary for 
few bright stars with unacceptable large errors in UCAC3). Therefore, we used a sample of 
stars with $UCACmag < 14$ and proper motion errors smaller than $10 mas/yr$ within the 
radius adopted for NGC~6167 and NGC~6193. To identify the members of the clusters we adopted 
the method developed by Cabrero-Ca\~{n}o \& Alfaro (1985). This statistical algorithm fits 
the observed proper-motion distribution with the sum of two Gaussian distributions, one 
elliptical for the field and one circular for the cluster. Moreover, we took into account 
the observed errors in the proper motions for individual stars, which was suggested by 
Zhao \& He (1990). The parameters of the two Gaussian distributions could then be determined 
from the maximum-likelihood technique. A summary of the application of these methods can be 
found in Fern\'andez Laj\'us et al. (2011). We note that the present study differs from that 
developed by Dias et al. (2006) who used a total sample of 416 stars extracted from the 
UCAC2 catalogue for the clusters under consideration.

Table~\ref{tab:pm} shows the results obtained for NGC~6167 and NGC~6193. Finally, 
Fig.~\ref{fig:vpds} presents the vector point diagram (VPD) for each cluster region. The 
adopted members (see Sect.~\ref{sec:optical}) produce  a clear clump separated from  the 
field stars.

\begin{table}
\caption{Kinematic results from brightest stars ($V < 14$) in NGC~6167 and NGC~6193 regions}
\centering
\begin{tabular}{cccc}
\hline
           & $Parameters$                 & $NGC~6167$ & $NGC~6193$ \\
\hline
           & $\mu_{\alpha\cos\delta_{c}}$ & -0.85      &  0.54      \\
$Cluster$  & $\mu_{\delta_{c}}$           & -2.61      & -2.91      \\
           & $\sigma$                     &  1.40      &  1.57      \\
\hline
           & $\mu_{\alpha\cos\delta_{f}}$ & -3.72      & -3.66      \\
           & $\mu_{\delta_{f}}$           & -3.85      & -6.16      \\
$Field$    & $\Sigma_1$                   &  8.34      &  6.76      \\
           & $\Sigma_2$                   &  8.03      &  9.12      \\
           & $\rho$                       &  0.24      &  0.28      \\
\hline
\label{tab:pm}
\end{tabular}
\begin{minipage}{7cm}
{\bf Note:} ($\mu_{\alpha\cos\delta_{c}},\mu_{\delta_{c}}$) and ($\mu_{\alpha\cos\delta_{f}},
\mu_{\delta_{f}}$) are the cluster and field proper motion centers in mas/yr;
$\sigma$ and ($\Sigma_1$, $\Sigma_2$) are the cluster and field proper motion
dispersions in mas/yr; $\rho$ is the obtained correlation coefficient.
\end{minipage}
\end{table}

\subsection{Optical photometric diagrams} \label{sec:optical}

\begin{figure*}
\begin{center}
\includegraphics[width=12cm]{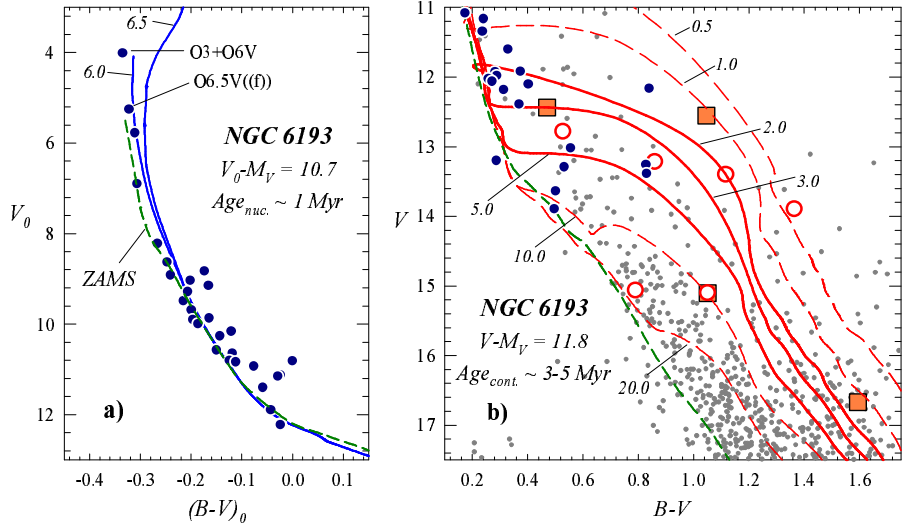}
\caption{Detailed CMD of stars located in NGC~6193 region. Symbols are the same as in Fig~\ref{fig:tcds1}.
  Dashed (green) curves are the Schmidt-Kaler (1982) empirical ZAMS; solid (blue and red) curves are
  Marigo et al. (2008) and Siess et al. (2000) isochones for $z = 0.02$. All the reference curves are
  corrected by the adopted colour excess value and apparent distance modulus (see Table~\ref{tab:param}).}
\label{fig:cmds1b}
\end{center}
\end{figure*}

\begin{figure*}
\centering
  \subfigure{\includegraphics[height=6cm]{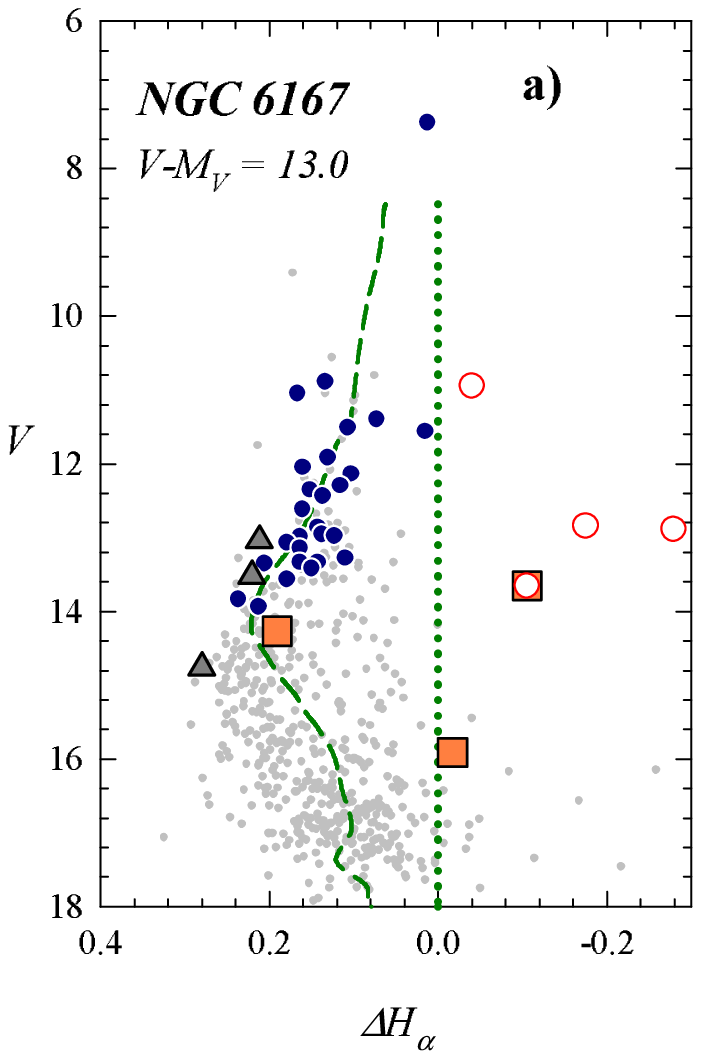}}
  \subfigure{\includegraphics[height=6cm]{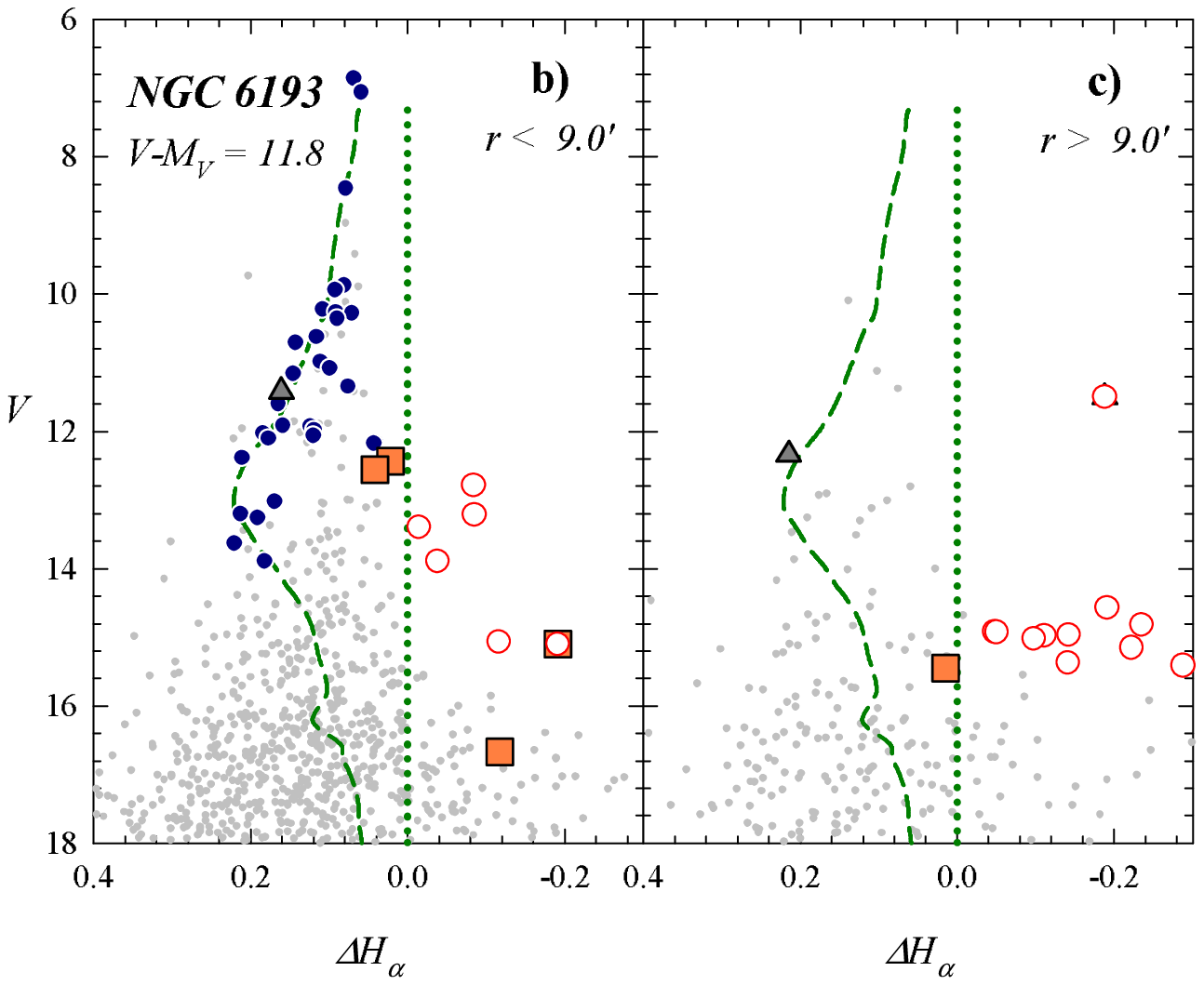}}
  \caption{$V$ vs. $\Delta~H_{\alpha}$ index of stars located in NGC~6167 region (left panel) and placed
  in NGC~6193 region and its surrounding (right panel). Symbols are the same as in Fig.~\ref{fig:tcds1}.
  Dashed (green) curves are the empirical MS form using Didelon (1982) and Feigelson (1983) parameters. 
  These curves are corrected by the adopted apparent distance moduli for each cluster (see
  Table~\ref{tab:param}). Dotted vertical lines indicate the adopted limit to separate $H_{\alpha}$ 
  emitting objects (see Sect.~\ref{sec:halpha}).}
  \label{fig:cmdha}
\end{figure*}

To perform the final membership assignment of the stars to the clusters, we analyze both: a) 
the individual stellar positions in all the  photometric diagrams (e.g. Baume et al. 2004, 
2006, 2009); and b) the obtained membership probabilities values computed from the kinematic 
information (see Sect.~\ref{sec:pmotion}). In the case of NGC~6193, $H_{\alpha(on/off)}$ 
data were also considered (see Sect.~\ref{sec:halpha}). The optical two-colour diagrams 
(TCDs) and colour-magnitude Diagrams (CMDs) of NGC~6167 and NGC~6193 are shown in 
Figs.~\ref{fig:tcds1} and \ref{fig:cmds1a}.

Individual positions of all stars down to $V = 14$ were simultaneously examined in all the 
diagrams. At fainter magnitudes, UCAC3 does not provide reliable solutions. This makes 
contamination by field stars more severe, preventing an easy identification of faint cluster 
members using this method.

The above selection of stars for both clusters allowed us to fit the blue edge of the 
adopted members with a properly reddened Schmidt-Kaler (1982) zero-age main-sequence (ZAMS) 
in the $U-B$ versus (vs.) $B-V$ diagrams. We adopted the well-known relation 
$E_{U-B}/E_{B-V} = 0.72 + 0.05~E_{B-V}$ and obtained the corresponding $E_{B-V}$ value for 
each cluster. The $B-V$ vs. $V-I$ diagram is routinely used as a diagnostic to detect 
deviations from the normal reddening law in a given direction. In this case the cluster 
diagrams suggest a normal law ($R = A_V/E_{B-V}$) for NGC~6167 and a marginally different
behavior for NGC~6193 confirming previous results of  Wolk et al. (2008b). 

We then simultaneously fitted a ZAMS or main sequence (MS) to all the CMDs to obtain cluster 
distance moduli, adopting $R = 3.1$ for NGC~6167 and $R = 3.2$ for NGC~6193. Finally, in 
Fig~\ref{fig:cmds1b}a we compared the observed upper MS of NGC~6193 with theoretical 
isochrones for solar metallicity, mass loss, and overshooting (Marigo et al. 2008). Some 
scatter (likely caused by the presence of binaries and rapid rotators) is present in our 
data and this prevents a unique isochrone solution. However, it is still possible to obtain 
an estimate of the nuclear age. All the resulting parameters are summarized in 
Table~\ref{tab:param}.

\begin{figure*}
\begin{center}
\includegraphics[width=16cm]{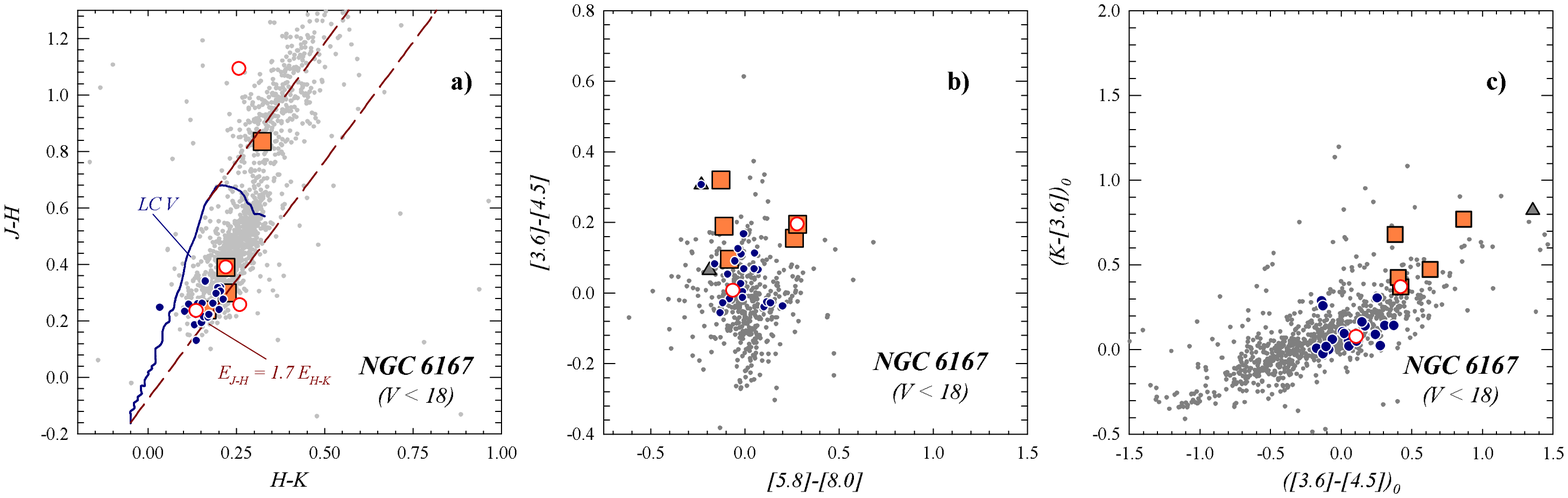}
\includegraphics[width=16cm]{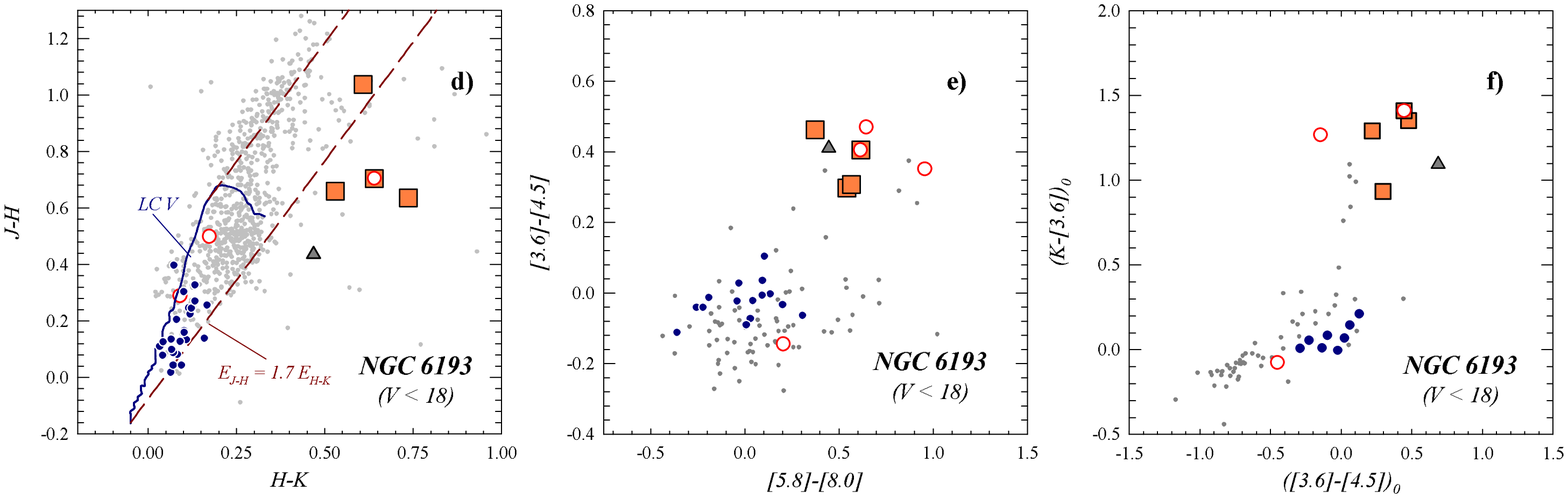}
\caption{Infrared CCDs of stars located in NGC~6167 (upper panels) and NGC~6193 (lower panels) regions.
  Symbols are the same as in Fig.~\ref{fig:tcds1}. Solid curves are the intrinsic Koornneef (1983)
  colours for MS stars.}
\label{fig:cmds1c}
\end{center}
\end{figure*}

\subsection{$H_{\alpha}$ photometric diagrams} \label{sec:halpha}

One of the aims of this work was to detect stars with $H_{\alpha}$ emission. For this
purpose we used the $\Delta H_{\alpha}$ index described in Sect.~\ref{sec:calib}. Similar 
indices were used in previous works (Adams et al. 1983; Sung et al. 1998; Baume et al. 2003).

Figure~\ref{fig:cmdha} presents the $V$ vs. $\Delta H_{\alpha}$ diagram for all stars with 
measured $\Delta H_{\alpha}$ index together with the corresponding ZAMS in this plane. Most 
stars follow approximately the ZAMS location indicating that they have no significant 
$H_{\alpha}$ emission allowing us to perform the calibration mentioned in 
sect.~\ref{sec:calib}. Therefore positive $\Delta H_{\alpha}$ index values correspond to 
absorption features and negative values to emission ones.

To detect stars with, on average, high $H_{\alpha}$ emission, we identified likely 
$H_{\alpha}$ emission objects with the brighest stars ($V < 15.5$) that had 
$\Delta H_{\alpha} - 3 e_{\Delta H{\alpha}}$ negative values.

As for the distribution of $H_{\alpha}$ emitters, we note in Fig.~\ref{fig:dss} an evident
concentration about $7\farcm0$ westward of NGC~6193. This concentration has a diameter
$\sim 2\arcmin$ and is located close to the NGC~6188 rim ($\alpha_{J2000} \sim 16:40:30$; 
$\delta_{J2000} \sim -48:44:20$; see Fig.~\ref{fig:dss}b), just at the edge of the SFO~80 
cloud (Sugitani et al. 1991; Sugitani \& Ogura 1994). This group of $H_{\alpha}$ emitters 
closely corresponds to  the MSX emission peaks at 8 $\mu$m found inside the cloud by 
Urquhart et al. (2009). All these elements seems to show different phases of stellar 
formation processes that are currently on-going. A particular spatial coincidence between a 
class II object (see Sect.~\ref{sec:ir}) and an $H_{\alpha}$ emitter was found very near to 
GSC~08329-03510. This star is placed just at the border of the HII region NGC~6188 and very 
close to the center of the concentration of $H_{\alpha}$ emitters 
(see Sect.~\ref{sec:halpha}).

\begin{table*}
\fontsize{8} {10pt}\selectfont
\caption{Parameters of the analyzed stellar groups}
\begin{tabular}{lcccccccccc}
\hline
Stellar group   & \multicolumn{2}{c}{Center}        & Radius     & $E_{B-V}$ & $R$ & $V-M_V$      & $A_V$     & $V_0 - M_V$  & \multicolumn{2}{c}{Age[Myr]} \\
                & $\alpha_{2000}$ & $\delta_{2000}$ & [']        &           &     &              &           &              & Nuclear & Contrac.           \\
\hline
NGC~6167        & 16:34:36.0      & -49:46:00.0     & $5.5$      & 0.80      & 3.1 & $13.0\pm0.2$ & $\sim2.5$ & $10.5\pm0.2$ & 20-30   & -                  \\
NGC~6193        & 16:41:24.0      & -48:46:00.0     & $9.0$      & 0.37      & 3.2 & $11.8\pm0.2$ & $\sim1.2$ & $10.7\pm0.2$ & $\sim1$ & $\sim3-5$          \\
IRAS~16375-4854 & 16:41:18.9      & -49:00:34.0     & $\sim2$    & $\sim2.2$ & 3.4 & $\sim18$     & $\sim7.5$ & $\sim10.5$   & $<1$    & -                  \\
\hline
\label{tab:param}
\end{tabular}
\begin{minipage}{12cm}
{\bf Note:} Radii adopted values to select the sources for analysis in each region.
\end{minipage}
\end{table*}

\subsection{Infrared photometric diagrams} \label{sec:ir}

Contracting stars emit an excess of radiation in the IR relative to their photospheric 
emission. This IR excess is considered to be produced by thermal emission from the 
circumstellar material. Therefore, this kind of objects are grouped into different 
''classes" according to their SEDs (Lada 1987; Andre et al. 1993). Thus, NIR and MIR 
photometric observations allow one to detect and distinguish this kind of sources. 

We adopted the selection scheme presented in Gutermuth et al. (2008). This method 
is based on the location of the observed objects in the $(K-[3.6])_0$ vs. $([3.6]-[4.5])_0$ 
plane using the standard dwarf star colours from Bessell \& Brett (1988) and Flaherty et al. 
(2007) colour excess ratios to obtain de-reddened colours. We then confirmed the selection by 
examining the locations in the $J-H$ vs. $H-K$ diagram (see Fig.~\ref{fig:cmds1c}).

This analysis allowed us to recognize class I and class II/III candidates among our data. 
Their spatial distribution (see Fig.~\ref{fig:dss}b) reveals that they are associated with 
IRAS sources and/or with denser molecular clouds reinforcing the hypothesis that they are 
indeed YSOs.

\section{Intrinsic cluster properties} \label{sec:objects}

\subsection{NGC~6167}

NGC~6167 appears to be an intermediate-age cluster with significant differential reddening
across its surface (see its TCDs and CMDs). This is consistent with the previous 
polarimetric results found by Waldhausen et al. (1999).

The age of the cluster was computed by fitting the distribution of the stars in $U-B$ and 
$B-V$ CMDs with isochrones. We paid particular attention to the brightest star in the CMDs, 
which is HD 149019 = CPD-49 9414 (A0 Ia star from $SIMBAD$). It has a high membership 
probability value from the kinematic analysis (see Sect.~\ref{sec:pmotion}), and is placed 
near the cluster center (see Fig.~\ref{fig:dss}c). The spectrophotometric method also infers 
for this star a distance moduli of $(V_0 - M_V)_{sp} = 10.5$, which is consistent with the 
adopted value for the cluster. As a consequence, this star is adopted as a likely cluster 
member and the best-fit isochrone is that for 20-30 Myrs. In addition, only few stars in 
the field of this cluster exhibit a sizeable $H_{\alpha}$ emission (see 
Fig.~\ref{fig:cmdha}a), bringing very support to this age value.

The second bright star in the cluster region is HD 149065 = CPD-49 9421 (B2 V star from 
$SIMBAD$). This star seems to be a foreground star according to its location in the 
photometric diagrams. In the $U-B$ vs. $B-V$ diagram, in particular, it shows a colour 
excess smaller than the cluster stars (0.24 and 0.80, respectively). Its spectrophotometric 
distance moduli is $(V_0 - M_V)_{sp} = 10.1$, which is significantly smaller than the one 
adopted for NGC~6167 (10.5). These differences might be caused by the Galactic absorption in 
the cluster direction (see Neckel \& Klare, 1980), which is systematically greater by $\sim 
2$ magnitudes (decreasing the mean brightness from less than 1 mag. to almost 3 mag.) at 
$\sim 1$ kpc from the Sun. On the other hand, HD 149065 is almost at the border of the 
adopted cluster region (see Fig.~\ref{fig:dss}c), so we can assume that it is probably an 
interloper.

With an age of 20-30 Myr, it is unlikely that the sources considered as class I/II objects 
in Fig.~\ref{fig:cmds1c} are real YSOs. These objects appear to be young based only on the 
MIR diagram (Fig.~\ref{fig:cmds1c}c), although their location in the NIR diagram 
(Fig.~\ref{fig:cmds1c}a) should be different to adopt them as YSOs.

\subsection{NGC~6193}

This cluster looks very young. Figure~\ref{fig:cmds1b}a presents the reddening-corrected CMD 
obtained for the adopted likely cluster members in the upper MS. Differential reddening was 
removed using individual colour excesses from Fig.~\ref{fig:tcds1}c and adopting $R = 3.2$. 
The isochrone fitting method (see Sect.~\ref{sec:optical}) then provides the  nuclear age 
($\sim$ 1 Myr). This result is compatible with the latest spectroscopic analysis of its 
brightest cluster member (star HD 150136; O3+O6V, Niemela \& Gamen 2005).

The lower MS of NGC~6193 shows quite  a significant $M_V$ scatter at any colour. There are 
also several objects in the cluster field, which were those identified as $H_{\alpha}$ 
emitters and a few as class II objects (see Sect.~\ref{sec:ir}). Figure~\ref{fig:cmds1b}b 
presents the location of these objects in a un-corrected CMD. These objects are all located 
above ($\sim$ 1-2 mag) the cluster ZAMS. Both effects, spread and location above the ZAMS, 
strengthen the idea that we are in the presence of a young cluster (cf. Fig. 3 in Preibisch \& 
Zinnecker 1999). The observed spread might have intrinsic causes such as 
binarity/multiplicity, the random distribution of the orientation of the accretion discs 
around single and multiple systems, as well as external causes such as differential 
reddening, field star contamination, and photometric errors.

The lower MS shape allowed us to derive an independent estimate of the cluster age (the 
contraction age) by comparing it with theoretical PMS isochrones. We achieved this by 
adopting the models computed by Siess et al. (2000). We note that it was impossible to find 
a unique isochrone that fitted the entire lower MS, but that we had to use a family of 
isochrone encompassing the stars loci in the CMD. Despite this, the mean location of the 
lower MS (see Baume et al. 2003) could be fitted with a group of isochrones that correspond 
to an age value marginally older than the nuclear age ($\sim$ 3-5 Myrs).

We are aware that the accuracy of the contraction age might only be considered as an order 
of magnitude. This is because the large dispersion of objects in the lower MS may also 
reflect a non-coeval process of stellar formation. This is reinforced by a) the presence of 
objects of different nature in that area of the CMD such as stars near the ZAMS, 
$H_{\alpha}$ emitters, and class II objects; and b) the existence of several young groups in 
the vicinity of NGC~6193 (see Wolk et al. 2008a,b and Fig.~\ref{fig:dss}b).

Nonetheless, in the region close to NGC~6193 there are only a few class I or class II 
objects, despite our identification of several likely sources as $H_{\alpha}$ emitters. Low 
mass sources in NGC~6193 would then be somewhat older than those associated with IRAS 
sources and molecular clouds. In addition, the small number of class I/II sources would 
indicated a small, if any, age dispersion in the cluster itself (see Fig.~\ref{fig:cmds1b}c).

\subsection{Embedded clusters}

\begin{figure}
\begin{center}
\includegraphics[width=8cm]{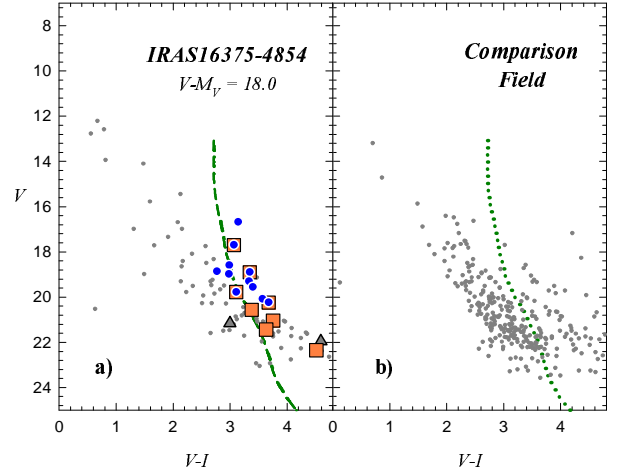}
\caption{$V$ vs. $V-I$ diagrams of stars located in IRAS~16375-4854 region
  and its corresponding comparison field. Filled (blue) circles are adopted
  members for each clump. Curves have the same meaning as in Fig~\ref{fig:tcds1}.}
\label{fig:cmds2}
\end{center}
\end{figure}

\begin{figure*}
\begin{center}
\includegraphics[width=16cm]{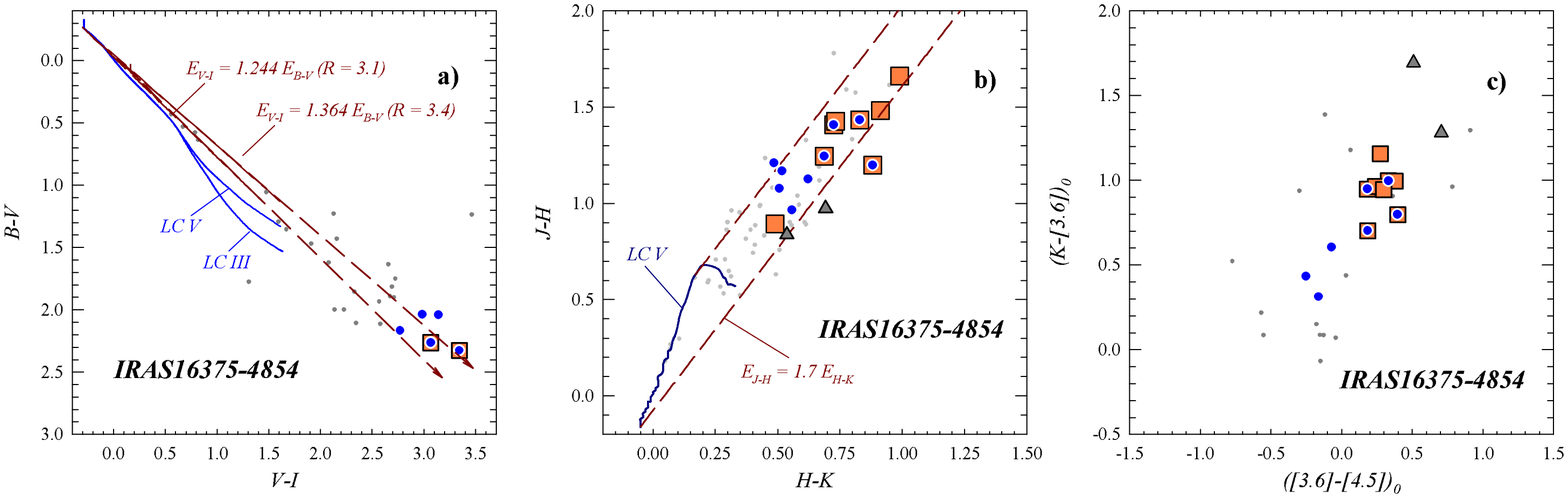}
\caption{NIR and MIR CCDs of stars located in IRAS~16375-4854 region. Symbols are the same as in
  Fig~\ref{fig:tcds1}. Solid curves are the intrinsic Koornneef (1983) colours for MS stars.}
\label{fig:tcds2}
\end{center}
\end{figure*}

\subsubsection{IRAS 16362-4845/RCW~108-IR}

This well known IRAS source contains the remarkable embedded and massive star formation 
region RCW~108-IR, which has been extensively studied in several wavelength ranges (from 
Radio to X ray; see Arnal et al. 2005, Comer\'on et al. 2005, Comer\'on \& Schneider 2007, 
Wolk et al. 2008a and references therein). Our observations cover nearly the entire region 
of this object. However, sources in our catalogues were selected based on their detection in 
the $V$ band and no further information to improve previous results could be achieved.

\subsubsection{IRAS 16375-4854 - IRAS 16379-4856}

These two IRAS sources are placed $\sim 15\arcmin$ southward of NGC~6193 and were partially 
covered by our optical survey. IRAS 16379-4856 has an associated CO peak (Arnal et al. 2003) 
whereas IRAS 16375-4854 is located in a diffuse optical nebula that has several associated 
X ray sources (Wolk et al. 2008a). Our reduced IRAC SST data indicate that both IRAS sources 
do not seem to be independent since the objects detected in these bands form a ring 
structure of about $10\arcmin$ in diameter, which is possibly a bubble centered at 
$\alpha_{J2000}\sim 16:41:37$; $\delta_{J2000} \sim -49:02:32$. This structure may have been
generated by some energetic event at that position.

By inspecting the stars detected on both regions, we note that IRAS 16375-4854 harbors 
objects brighter than IRAS 16379-4856. By using the corresponding optical photometric 
diagrams (CMDs and TCDs; see Figs.~\ref{fig:cmds2} and \ref{fig:tcds2}a), we estimated its 
main parameters (see Table~\ref{tab:param}). They were obtained by selecting the most 
probable members checking simultaneously their individual position on all the photometric 
diagrams and also the finding chart. Apparently this region is affected by a non-ISM 
interstellar reddening ($R \sim 3.4$) and has a distance similar to NGC~6193 and RCW~108-IR.

As for the infrared photometric diagrams of IRAS 16375-4854, the position of the selected 
sources as YSOs appears to be unlikely. They might be class II/III objects, although we need 
to analyze X-ray observations to confirm this assumption.

\section{Discussion and conclusions} \label{sec:discussion}

Previous investigations by our group (V\'azquez et al. 1996; Baume et al. 2003) presented 
optical observations (typically $UBVRI$) of young star clusters and star forming regions. To 
reduce the uncertainties, we complement these $UBVRI$ data with $H_{\alpha}$ photometry, 
kinematic information, and near/mid infrared data. This yields more reliable values of 
reddening, distance, and age for NGC~6167, NGC~6193, and the IRAS~16375-4854 source. As a 
consequence, we can establish that all these groups are placed at the same distance from the 
Sun ($\sim$ 1300 pc) in the Sagittarius-Carina Galactic arm, but that their ages differ 
widely (see Table 4). We find that:
\begin{itemize}
 \item NGC~6167 is found to be an intermediate-age cluster ($\sim$ 20-30 Myr);
 \item NGC~6193 is very young ($\sim$ 1-2 Myr) with PMS objects at different stages, 
       $H_{\alpha}$ emitters, and class II objects;
 \item Finally, IRAS~16375-4854 is the youngest of the three and contains several YSOs.
\end{itemize}

In the specific case of NGC~6193, PMS stars appear to be slightly older than MS stars, i.e.
the contraction age is older than the nuclear age, which might be an indication of 
non-coevality.

The main result of this work is the homogeneity of distances, and the differences between 
the ages of the Ara OB1a clusters, which may place on firmer ground future studies of the 
interactions between the different stellar and interstellar components of the region, such 
as those suggested by Arnal et al. (1987). The results presented here support the picture in which Ara OB1a is a region where star formation has 
proceeded for several tens of millions of years up to the present.

\begin{acknowledgements}
GB acknowledges ESO for granting a visitorship at Vitacura premises in March 2010, where 
most of this work was done. He also acknowledges the support from CONICET (PIP~5970) for the 
trip to CTIO on March 2006, where part of the data have been taken. We thanks very much the 
staff of the observatories CTIO, LCO and CASLEO during all the runs related with this work. 
The authors are much obliged for the use of the NASA Astrophysics Data System, of the 
$SIMBAD$ database and $ALADIN$ tools (Centre de Donn\'es Stellaires ---Strasbourg, France) 
and of the WEBDA open cluster database. This publication also made use of data from the 
2MASS, which is a project of the University of Massachusetts and the Infrared Processing and 
Analysis Center/California Institute of Technology, funded by the NASA and the NSF. 
We thank the Cambridge Astronomical Survey Unit (CASU) for processing the VISTA raw 
data. This work is based (in part) on observations made with the SST, which is operated by 
the JPL, California Institute of Technology under a contract with NASA. We also thank the 
referee, whose comments helped to improve the paper significantly.
\end{acknowledgements}

\end{document}